\documentclass[]{spie}  
\usepackage[]{graphicx}
\usepackage{mathtools}

\title{Evolved stars at high angular resolution: present and future} 

\author{Claudia Paladini\supit{a}
\skiplinehalf
\supit{a}Institut dÕAstronomie et dÕAstrophysique, Universit\'e libre de Bruxelles, CP 226, Boulevard du Triomphe, 1050 Bruxelles, Belgium;\\
}

\authorinfo{For further information e-mail to paladini@ulb.ac.be\\ }

\begin{document} 
  \maketitle 

\begin{abstract}
The late evolutionary stages of stellar evolution are a key ingredient for our understanding in many fields of astrophysics, including stellar evolution and the enrichment of the interstellar medium (ISM) via stellar yields. Already the first interferometric campaigns identified evolved stars as the primary targets because of their extended and partially optically thin atmospheres, and the brightness in the infrared. Interferometric studies spanning different wavelength ranges, from visual to mid-infrared, have greatly increased our knowledge of the complex atmospheres of these objects where different dynamic processes are at play. In less than two decades this technique went from measuring simple diameters to produce the first images of stellar surfaces. By scanning the extended atmospheres we constrained theoretical models, learnt about molecular stratification, dust formation, and stellar winds, and there is still a lot to be done. In this contribution I will review the recent results that optical/infrared interferometry has made on our current understanding of cool evolved stars. The presentation will focus on asymptotic giant branch stars, and red supergiants. I will discuss the challenges of image reconstruction, and highlight how this field of research will benefit from the synergy of the current interferometric instrument(s) with the second generation VLTI facilities GRAVITY and MATISSE. Finally I will conclude with a short introspection on applications of a visible interferometer and of the the Planet Formation Imager (PFI) to the field of evolved stars.
\end{abstract}



\section{INTRODUCTION}
\label{sec:intro} 
%
The stars covered in this review are the so-called asymptotic giant branch (AGB) and red supergiant stars (RSG).
It is general knowledge that the initial mass on the main sequence will determine the fate of a star.
Objects with a mass in the range 1-8~$M_{\odot}$, after going through the red giant and horizontal branch phase, will become AGB stars.
During this phase, stars will expel their envelope via stellar wind, and end their lives as white dwarf.
Stars more massive than 8-10~$M_{\odot}$ will have an almost horizontal excursion across the Hertzsprung-Russels diagram. After a quick phase as yellow supergiant 
they will become RSG, ending their lives with supernovae explosion.
AGB and RSG stars have bolometric luminosity up to $10^4~L_{\odot}$. The cool rarefied atmospheres have temperature in the range of 2\,000 - 4\,000~K
and are the favoured place for the formation of various molecular and dust species.
Both AGB and RSG stars are known to be a crucial player in the cosmic circuit of matter.
Through the mass-loss process, these stars return the elements synthesised during their life to the interstellar medium. This material
acts as building block for the next generation of stars, and mass loss is one of the main (still unknown) ingredients in stellar evolutionary codes.

Investigated since the early '70, the mass-loss process from evolved stars is still matter of discussion in the scientific community.
Several open questions include: i) what drives the stellar wind in oxygen-rich AGB stellar atmospheres? ii) what drives the wind for the 
low-pulsating (or less evolved) AGBs? iii) what drives the wind in RSG? iv) what are the seeds for (oxygen-rich) dust formation; 
v) how does the stellar wind affects the various spatial scales of 
the stellar atmosphere (from the photosphere to the interface with the stellar envelope). 
This review will summarise where are we standing with the latter question. 

I will focus on the ``2-dimensional'' information obtained over the last few years 
on the inner spatial scales probed with long baseline interferometry. With 2-dimensional information, I mean the possible asymmetric structures that
have been detected  so far (via visibility model-fitting, direct closure phase and differential phase measurements), and their interpretation.
Interested readers may want to consult also earlier reviews on the topic [\citenum{scholz2003},\citenum{perrin2003},\citenum{wittkowski2014}].
Results obtained with other interferometric techniques such as aperture masking or speckle interferometry 
are not covered here. I encourage the readers to look up very interesting recent results 
reported by Tuthill in the current proceeding (contribution 9907-13).

The review is divided in 4 main sections. The first part is dedicated to the studies conducted in the near-IR, where 
the photosphere and inner molecular envelope of giant stars (1-5 stellar radii) are observed.
The second section will focus on the mid-IR studies, where dust formation occurs (up to $\sim$ 10 stellar radii).
The near-future prospects as well as more long-term forecasts will be given in the Section~4, followed by conclusing remarks 
in Section 5.   
\section{Photosphere and inner molecular environment} \label{near-ir.sect}

Departure from circular symmetry has been known for giant stars from various high angular resolution observations [\citenum{ragland2006}, and references therein].
For many years the interpretation of such structures was not univocal and involved (i) spots due to convection, or (ii) elliptical distortion due to stellar rotation [\citenum{vanbelle2013} and  \citenum{cruzalebes2015} are very recent examples].
However, at least in the case of AGB stars, high stellar rotation rates are usually connected to the presence of a binary companion transferring momentum to the primary 
[\citenum{barnbaum1995}, \citenum{mayer2014}]. 
So far, convection appears to be the most likely explanation for the deviation from symmetry.

Despite what is observed on the Sun, where at least $10^7$ granules populate the surface,
Schwarzschild [\citenum{schwarzschild1975}] predicted the presence of only few convective cells
on the stellar surface of AGB and RSG stars. In fact, a star evolving off the main sequence, 
inflates its radius and lower its gravity. Gravity together with temperature are the main parameters controlling the convection.
When the star reaches the red giant branch the number of granules drops to a few thousand. 
The stellar surface becomes more and more irregular, and by the time the star is a RSG or an AGB, only a few
convective patterns are expected to populate the stellar surface [\citenum{chiavassa2015}].
As soon as the first images of RSG stars were available, the dichotomy between ellipticity vs. spots was solved.
In [\citenum{haubois2009}] the stellar surface of Betelgeuse was imaged with IOTA. The images are reconstructed using WISARD [\citenum{meimon2005}] and 
MiRA [\citenum{thiebaut2008}] algorythms. A model prior and some regularization conditions are involved in the image reconstruction, but the two spots were 
confirmed via geometric model fitting of the visibilities. The authors mention that in analogy to what is observed in the Sun,
the bright spots could be faculae, i.e. bright zones surrounded by dark convection cells. Connection with magnetic field is suggested, but could not be 
confirmed by these solo observations.
Finally the same set of data was compared with 3D model predictions in [\citenum{chiavassa2010b}] who confirmed the convective nature of the structures,
and identified H$_2$O as main opacity source in the image.
Over the last years we have seen a few other cases of RSG images, and all of them had in common the detection of at least
one large convective spot [\citenum{chiavassa2010a}, \citenum{baron2014}, \citenum{monnier2014}]. 

The first data sets allowing image reconstruction for AGB stars showed more or less asymmetric (molecular) shells [\citenum{ragland2008}, \citenum{lacour2009}, \citenum{lebouquin2009}, \citenum{haubois2015}]
with wavelength- and time-dependent global sizes.
No obvious evidence of convective patterns are detected, however this is mainly the result of non optimal uv-coverage/visibility curve sampling. The recent image of R~Car obtained from PIONIER data [\citenum{monnier2014}] confirms that spots are also detected on the surface of AGBs. At the time when this review is being written, at least three more stars have been imaged with PIONIER. Papers are in preparation [Paladini et al., Wittkowski et al.] and confirm the tendency of surface/convection related structures also in AGB stars. Preliminary results seems to support the idea that in the near-IR carbon-rich AGB stars are more asymmetric than their oxygen-rich counterpart [\citenum{cruzalebes2015}]. 

Ascertained that convection is a crucial ingredient in the modelling of AGB and RSG, now one needs to determine: (i) what are the time-scales of convective patterns; (ii) how does the convection interplay with pulsation; (iii) what is the role of magnetic field. While 3D models are still under development [\citenum{freytag2008}, \citenum{freytag2015}, \citenum{chiavassa2015}], observational astronomers, and more precisely optical interferometrists, should push to obtain more images, and also to compare the observations with 1D available models. 
A summary of what has been done with 1D models is available in [\citenum{wittkowski2014}], nevertheless I would like to mention here a few interesting developments occurred in the field of RSG over the last few years. Usually 1D models successfully reproduce spectroscopy and photometric data. However this is not always the case for interferometric measurements, especially when it comes to RSG. Asymmetric structures can explain the disagreement only partially, as they mainly show up at high spatial frequencies.
In a series of works [\citenum{arroyo-torres2013}, \citenum{arroyo-torres2014}, \citenum{arroyo-torres2015}] showed that PHOENIX hydrostatic models are too compact to reproduce the observed extension of RSG stars in the near-IR, while on the other hand they can reproduce the observations of low-pulsating AGBs. The use of 3D models including convection prescription does not improve the situation of RSG, and neither the 1D CODEX models implementing pulsation.
A correlation between visibility ratio (continuum / CO (2-0) bandhead) of RSGs and the luminosity and surface gravity is observed. Extension of the atmospheres of RSG and mira stars are comparable [\citenum{arroyo-torres2015}]. However the correlation between luminosity and surface gravity just mentioned is not observed in the case of mira variables. This points towards
the possibility that the extension of the layers of RSG is not triggered by shocks generated during pulsation, as in the case of miras. The authors suggest that radiatively driven extension caused by radiation pressure on Doppler-shifted molecular lines might help to levitate the layers. As also magnetic field has been measured on the surface of RSG [\citenum{auriere2010}],
it might very well be that in the end all these mechanisms have to be taken into account during modelling.
Within this frame, simultaneous interferometric and polarimetric observations will help to clarify what is the interplay between magnetic field and convection [\citenum{montarges2016}, \citenum{auriere2016}].

\section{The dust-forming region}
Both AGB and RSG show infrared dust excess in their spectra. The details of dust formation and acceleration process are matter of discussion [Sect.~3, \citenum{wittkowski2014}].
Departure from spherical symmetry have been observed also in the dust forming region ($\sim10~\mu$m), at the onset of stellar wind.
Asymmetric structures around AGB stars detected have been interpreted in terms of molecular or dust blobs [\citenum{chandler2007}, \citenum{paladini2012}, \citenum{sacuto2013}], circumbinary/circum-companion discs [\citenum{deroo2007}, \citenum{ohnaka2008b}, \citenum{klotz2012}], or as the signature of a binary companion [R~Aqr, \citenum{tatebe2006}].
The picture of RSG is similar to the one on the AGB: [\citenum{ravi2011}] interpreted asymmetric structures observed around Betelgeuse as related to the photospheric convective patterns,
while [\citenum{ohnaka2008a}] reported the presence of a pole-on torus around WOH G64.
Although additional observations obtained with single-dish telescopes support the fact that both dust-plumes and torus are present around AGB and RSG [\citenum{dewit2008}, \citenum{kervella2009}, \citenum{kervella2011}, \citenum{ohnaka2014}], it is obvious that there is no wide consensus on what really causes the departure from symmetry observed in the interferometric data. 
This is mainly due to the lack of imaging capabilities of the current (and past) interferometers in this wavelength range, 
but the scenario will improve as soon as VLTI/MATISSE will be on sky. 

\section{Outlook}

Since the early days of long baseline interferometry, AGB and RSG  have been favourite targets for their high luminosity and extended atmospheres. 
Imaging is relatively challenging (see the contributions of Baron, Soulez, and Sanchez in this proceeding). 
The very complex environments and time variability 
require a good uv-coverage and sampling of the visibility curve, and the data cannot be combined from one year to the other. 
Observations should be collected and combined within one month for the most variable objects (i.e., mira stars). 
Current instrumentations with 4- and 6-telescopes have shown that images can be done more or less routinely in the near-IR, but so far
we were limited to few spectral channels or broad band.

GRAVITY, the second generation instrument recently installed at VLTI, will provide high spatial and spectral resolution.
Spectra at each position of the stellar surface across the $K$-band will be available. CO, H$_2$O, OH, HCN, C$_2$, and C$_2$H$_2$
molecules will be mapped as well as the continuum. If with PIONIER and MIRC we can already image convective patterns, 
the spectral resolution will allow to localise and characterise these structures in more detail [\citenum{ohnaka2011}, \citenum{ohnaka2013}, \citenum{ohnaka2016}]. 
Time-series will allow to follow the evolution of such pattern and also the interplay with pulsation.
The possibility of obtaining a simultaneous polarization information is quite intriguing, although very challenging according to the experts.
Polarization measurements combined with the high spatial and spectral resolution would allow to clarify how much magnetic field plays a role
in the convection theory for this class of stars.

MATISSE will open new spectral windows in the $L$ and $M$ bands, and will deliver images also in the $N$-band allowing to
solve the dichotomy discs vs. blobs. Very extended targets such as the nearby AGBs and most of the RSG might well
benefit from the aperture masking experiments available on NACO, SPHERE, and VISIR to complement and complete
the visibility data sets of long baseline interferometry. 
Since a connection between apparent surface features and the morphology of the dust shell has not been established yet,
and given the variability of the stars, we would benefit from simultaneous observations. 
In this frame an experiment able to combine the various wavelength ranges (PIONIER+GRAVITY+MATISSE)
 and collect them in one shot would be very welcome (so-called i-Shooter interferometer).
 
 On a longer time-scale, an extension of the observing window to the visible will allow to probe the deeper layers of these giant stars,
 and to study the hot chromospheric plasma. However, to be successfull such a visible instrument will have to take into account that cool giant stars 
 might be as faint as 7-13 mag in $V$-band.
Long baselines of the order of km foreseen by PFI [Kraus et al., this proceeding] in the thermal infrared will allow a breakthrough in the field.
Galactic globular clusters, and possibly the Magellanic Clouds might be easily reached. The interferometric community will have the chance 
in such case to give a strong contribution to questions such as how the mass-loss process depends on metallicity.

\section{Concluding remarks}

In this contribution I briefly summarised the progress that long baselines interferometric studies have brought
in the field of cool evolved stars, with a focus on the geometry of the environment. 
Time series of images at high spectral resolution for a benchmark of objects are needed to constrain the theory.
Such kind of study needs to be coordinated with other multi-wavelengths and multi-technique observations in order to be able
to tackle the main questions highlighted in the text. Given the complexity and the variable nature of these stars,
time should be invested to investigate a polarization mode, and also to retrieve flux calibrated spectra/photometry 
from the current interferometric facilities.

\acknowledgments     
PC is supported by the Belgian Fund for Scientific Research F.R.S.- FNRS.
I would like to acknowledge all my colleagues from the cool stars and interferometry community
for many interesting discussions and crazy ideas. 

\bibliography{evolbib}   
\bibliographystyle{spiebib}   

\end{document}